\documentclass[aps,prd,eqsecnum,amsmath,amssymb,longbibliography,10pt,notitlepage]{revtex4-1}
\usepackage[utf8]{inputenc}
\usepackage[colorlinks=true, linkcolor=blue, citecolor=teal]{hyperref}
\usepackage{bm}
\usepackage{latexsym,amssymb,amsmath,amsfonts}
\usepackage{graphicx}
\usepackage{longtable}
\usepackage{physics}
\usepackage{bm}
\usepackage{footnotebackref}
\usepackage{multirow}
\usepackage[usenames]{xcolor}
\usepackage{soul}
\usepackage{makecell}
\usepackage[flushleft]{threeparttable}
\usepackage{graphicx}
\usepackage{amssymb}
\usepackage{longtable}
\usepackage{rotating}
\usepackage{color}
\usepackage{ifthen}
\usepackage{slashed}
\usepackage{xspace}
\usepackage{tabularx}
\usepackage{float}
\usepackage{dutchcal}

\newcommand\apjl{Astrophysical Journal Letters}

\def \hatOm {{\hat{\bm{\Omega}}}}
\def \hatp {{\hat{\bm{p}}}}
\def \d {{\rm d}}
\def \be{\begin{equation}}
\def \ee{\end{equation}}

\begin{document}

\title{Cosmic variance of the Hellings and Downs correlation for ensembles of \\ universes having nonzero angular power spectra}

\author{Deepali Agarwal}
\email{deepali.agarwal@uclouvain.be}
\affiliation{Centre for Cosmology, Particle Physics and Phenomenology (CP3), Universite catholique de Louvain, Louvain-la-Neuve, B-1348, Belgium}
\affiliation{Inter-University Centre for Astronomy and Astrophysics (IUCAA), Pune 411007, India}

\author{Joseph D. Romano}
\email{joseph.romano@utrgv.edu}
\affiliation{Department of Physics and Astronomy, University of Texas Rio Grande Valley, One West University Boulevard, Brownsville, TX 78520, USA}

\begin{abstract}
Gravitational waves (GWs) induce correlated perturbations to the arrival times of pulses from an array of galactic millisecond pulsars.
The expected correlations, obtained by averaging over many pairs of pulsars having the same angular separation ({\em pulsar averaging}) and over an ensemble of model universes ({\em ensemble averaging}), are described by the Hellings and Downs curve. As shown by Allen [Phys. Rev. D 107, 043018 (2023)], the pulsar-averaged correlation will not agree exactly with the expected Hellings and Downs prediction if the gravitational-wave sources interfere with one another, differing instead by a ``cosmic variance" contribution.
The precise shape and size of the cosmic variance depends on the statistical properties of the ensemble of universes used to model the background. 
Here, we extend the calculations of the cosmic variance for the standard Gaussian ensemble to an ensemble of model universes which collectively has rotationally invariant correlations in the GW power on different angular scales (described by an {\em angular power spectrum}, $C_l$ for $l=0,1,\cdots$.). 
We obtain an analytic form for the cosmic variance in terms of the $C_l$'s and show that for realistic values $C_{l}/C_0\lesssim 10^{-3}$, there is virtually no difference in the cosmic variance compared to that for the standard Gaussian ensemble (which has a zero angular power spectrum).
\end{abstract}

\maketitle

\section{Introduction}

Recent observations taken by four worldwide pulsar timing array (PTA) collaborations have shown weak to compelling evidence of a low frequency ($\sim 10^{-9}$ Hz) correlated gravitational-wave (GW) signal~\cite{Reardon_2023,Xu_2023,EPTA_InPTA_23,Agazie_2023}. These observations have opened up a new window to gravitational-wave astrophysics. For example, it potentially sheds light on exotic physics such as cosmic inflation, cosmic strings and dark matter in the Milky Way, etc.~\cite{Afzal_2023}, and provides an important test bed for alternative theories of gravity~\cite{Arzoumanian_2021, NG15yr-altpol}. The International Pulsar Timing Array’s (IPTA)~\cite{Hobbs_2010} third data release is currently under preparation. This dataset will combine the 15-yr dataset from the North American Nanohertz Observatory for Gravitational Waves (NANOGrav) with the latest datasets from the Eurpoean~\cite{Kramer_2013}, Indian~\cite{Joshi2018}, and Parkes~\cite{Hobbs_2013} PTAs. This combined dataset is expected to have better sensitivity due to an increased number of pulsars and a longer effective observation time. 

PTAs employ arrays of galactic millisecond pulsars, which are highly stable clocks. The observed pulsars emit beams of radio waves, which intersect our line-of-sight every rotational period. The arrival times of the pulses have been observed for years to decades, and a timing model is fit for each pulsar. When a GW passes between the Earth and a pulsar, the arrival times deviate from the expected values. The timing residuals or redshift of radio pulses from pairs of pulsars are correlated to look for evidence of the common underlying GW signal. The correlation $\rho_{12}$ of the redshifts $z_1$, $z_2$ observed in a pair of pulsars is defined as the time-averaged product
\begin{equation}\label{eq:def_Corr}
\rho_{12}\equiv\overline{z_1(t)z_2(t)}\equiv \frac{1}{T}\int_{-T/2}^{T/2}{\rm d}t\> z_1(t) z_2(t)\,,
\end{equation}
where $T$ is the total observation time. The subscripts ``1" and ``2" label the two pulsars. The correlation leads to an expected pattern that depends upon the Earth-source-pulsar geometry.

The expected pattern is a quasiquadrupolar curve for an unpolarized, isotropic stochastic GW background (SGWB), depending only on the separation angle between a pair of pulsars. This pattern is known as the Hellings-Downs (HD) curve~\cite{Hellings1983}. For an anisotropic background, this pattern has different shapes for the different spherical harmonic components of the anisotropy~\cite[Fig.~2]{Mingarelli2013}, as observed by a pair of pulsars in the $xz$ plane. 
But if we average the correlation over many pairs of pulsars with the same angular separation, one again recovers the HD curve~\cite{Cornish_2013}.

The PTA experiments use the HD curve as a template to detect and infer the source properties. Recent findings by~\citet{Allen23} have deepened our understanding of the HD curve, and a few important facts are as follows:
\begin{itemize}

\item The HD curve denotes the expected correlation between a pair of pulsars separated by angle $\gamma$. 
In their seminal paper~\cite{Hellings1983}, Hellings and Downs obtained this curve by averaging the correlations for a fixed pair of pulsars over an isotropic distribution of GW sources on the sky, assuming that the GW sources did not interfere with one another.

\item Thirty years later, Cornish and Sesana~\cite{Cornish_2013} showed that the HD curve also arises for a single GW point source (deterministic sinusoidal signal), provided the correlation is averaged over pulsar directions, keeping the separation angle fixed, i.e., {\em pulsar averaging}, $\langle\rho_{12}\rangle_{12\in\gamma}$.

\item In the absence of noise, deviations of the measured correlation away from its expected value have two contributions: {\em pulsar variance} and {\em cosmic variance}. Pulsar variance is due to the differing response of pairs of pulsars pointing in different directions on the sky, but having the same angular separation. Although pulsar variance can be reduced by binning and averaging the correlations of pulsar pairs having (approximately) the same angular separation, the cosmic variance due to interfering GW sources cannot be removed; it can only be inferred.
\end{itemize}
Pulsar and cosmic variance due to an isotropic background are also discussed in~\cite{Bernardo_2022,Bernardo_2023, calıckan2023} using the harmonic-space formalism~\cite{Roebber_2017}. 

Although previous works~\cite{Cornish_2013,PhysRevD.95.042002} have suggested that deviations from the standard Gaussian ensemble would not lead to significant deviations away from the HD mean, \citet{Allen23} argued that the second moment of the pulsar-averaged correlations (i.e., the cosmic variance) also carries important information about the statistical nature of the ensemble from which our Universe is a single realization. In this paper, we analytically determine how nontrivial rotationally invariant correlations in the GW power on different angular scales affect the cosmic variance. 
Such correlations are typically described by an {\em angular power spectrum}, $C_l$ for $l=0,1,\cdots$ 
[see \eqref{eq:def_statsIso_pix} and \eqref{eq:Cpix2CSph}]. These investigations will help us assess if the deviation of the observations from the HD curve can be attributed to such correlations.

For example, supermassive black-hole binaries (SMBHBs) in the centers of merging galaxies are the most natural source of the GWs to be observed by PTA experiments, and the local unresolved SMBHBs can introduce a departure from the predictions for a standard Gaussian ensemble due to structure in the local universe~\cite{Cornish_2013,PhysRevD.102.084039,Mingarelli2017}. The reason is as follows: the sky distribution of the SMBHB population is expected to follow the galaxy distribution, which is anisotropic, and angular correlations are observed in galaxies embedded in cosmological large-scale structure~\cite{10.1093/mnras/stx721,10.1093/mnras/stu2693,VIPERS-2017}. Predictions for the expected anisotropies and angular correlations in the intensity distribution of the GW background in the LIGO frequency band have also been explored~\cite{PhysRevD.96.103019,Jenkins2018,PhysRevD.98.063501,PhysRevLett.120.231101,PhysRevD.101.103513}. A search for angular correlations in the GW power was performed by the NANOGrav collaboration using their 15-yr dataset~\cite{NanogravAniso2023}, but it did not find any significant evidence for a nonzero angular power spectrum. However, the frequentist analysis performed in that work did not account for cosmic variance~\cite{Allen23,Bernardo_2022,Bernardo_2023} and inter-pulsar-pair covariance~\cite{PhysRevD.103.063027,AllenRomano23}. 

The rest of the article is structured as follows: We start in Sec.~\ref{sec:timingResponse} with a brief introduction to the response of a PTA to GWs. Section~\ref{sec:SGWB} provides the approach used to describe ensembles of GW universes having nonzero angular power spectra using a two-stage ensemble averaging process~\footnote{Such a two-stage averaging process was first explicitly described in \cite[Fig.~2]{Jenkins-et-al:2019}, although incorrectly assuming that the different realizations of GW power on the sky were Gaussian distributed.}. 
An expression for the cosmic variance for such ensembles is derived in Sec.~\ref{sec:CV_Position}. We conclude in Sec.~\ref{sec:Conclusions} with a brief summary and possible future directions. 

\section{Timing Residual Response to Gravitational Waves} \label{sec:timingResponse}

The metric perturbations at a space-time point $(t,\vec{x})$ can be written as a sum of plane waves in the synchronous transverse-traceless gauge as
\begin{equation}\label{eq:waveExpansion}
\begin{aligned}
    h_{ij} (t,\vec{x})= \sum_{A=+,\times} &\int_{-\infty}^{\infty} \d f\int_{S^2}\d^2\hatOm\,\tilde{h}_A(f,\hatOm)\,e^A_{ij}(\hatOm)\,e^{i2\pi f(t-\vec{x}\cdot\hatOm/c)}\,.
\end{aligned}
\end{equation}
Here, $\tilde{h}_A(f,\hatOm)$ is the Fourier domain component of the metric perturbations having frequency $f$, propagation direction $\hatOm$, and polarization $e^A_{ij}(\hatOm)$. The timing residuals depend upon the integrated projection of the metric perturbations along the path traveled by a radio pulse from the pulsar to Earth. The timing residual evaluated at time $t$ for a single pulsar in direction $\hat p$ at a distance $D$ from Earth is given by~\cite{romano2023answers}
\begin{equation}
    \Delta T(t) = \frac{1}{2c} \hat p^i \hat p^j \int_{s=0}^{D}\,\d s\, h_{ij} [\tau(s),\vec{x}(s)]\,,
\end{equation}
where the space-time path between the pulsar and the Earth is parameterized as
\be
\tau(s)=t - (D-s)/c\,,
\qquad
\vec x(s)=(D - s)\hat p+\vec{r}_2\,,
\ee
 where $\vec r_2$ is the position vector of the Earth at the time $t$ when the observation is made. Note that we can use this straight line path since the metric perturbations are already first-order small.

The redshift $z(t)$ is related to the timing residual via differentiation with respect to $t$:
\begin{equation}
z(t) \equiv \frac{{\rm d}\Delta T(t)}{{\rm d}t}= \frac{1}{2c} \hat p^i \hat p^j \int_{s=0}^{D}\,\d s\, \frac{\partial h_{ij}}{\partial t} [\tau(s),\vec{x}(s)]\,.
\end{equation}
Performing the integration, we find~\cite{romano2023answers}
\begin{equation}\label{eq:redshift}
\begin{aligned}
z(t)=\sum_{A=+,\times} &\int_{-\infty}^{\infty} \d f\int_{S^2}\d ^2\hatOm\,\tilde{h}_A(f,\hatOm)\,R^A(f,\hatOm, D\hatp)\,e^{i2\pi f(t-\vec{r}_2\cdot\hatOm/c)}\,,
\end{aligned}
\end{equation}
where $R^A(f,\hatOm,D\hatp)$ denotes the {\it redshift} pulsar response~\cite{romano2023answers}:
\begin{equation}
    R^A(f,\hatOm, D\hatp)\equiv F^A(\hatOm, \hatp)\left[1-e^{-\frac{2\pi i fD(1+\hatOm\cdot\hat{p})}{c}}\right]\,,
    \qquad
    F^A(\hatOm, \hatp)\equiv \frac{1}{2} \frac{\hat p^i \hat p^j }{1+\hatOm\cdot\hatp}\,e^A_{ij}(\hatOm)\,.
\end{equation}
The terms in square brackets are called the ``Earth" and ``pulsar" terms, respectively. Being highly oscillatory relative to $F^A(\hatOm, \hatp)$, the pulsar term can be ignored in most circumstances when correlating data from multiple pulsars, as discussed in~\citet{Allen23}. Since this is also the case for pulsar averaging, which we use to obtain the cosmic variance, most of the key expressions that follow will involve the Earth-term-only response function $F^A(\hatOm,\hatp)$ for the redshift. The vector $\vec{r}_2$, which is the position vector of Earth in the Solar System barycenter frame, can also be set to zero relative to the wavelength of the GWs that PTAs are sensitive to (of order 10's of light-years).

The redshift (\ref{eq:redshift}) can be evaluated for any GW source. In the next section, we discuss a signal model that describes a rotationally invariant ensemble of universes, which have nontrivial correlations in the GW power on the sky.

\section{Characterization of rotationally invariant correlations in GW power}\label{sec:SGWB}

The signal associated with a GW background may be either deterministic or stochastic depending on the sources that produce it. 
Some relevant parameters that determine this outcome include the number and location of the sources on the sky, their frequency evolution, and their amplitude relative to competing instrumental and environmental noise. In the absence of definitive knowledge of these parameters, the best we can do is create ensembles of model universes against which we can compare predictions to the actual observations.

The standard Gaussian ensemble is one such ensemble of GW universes.  
Each universe in this ensemble is anisotropic, but the collection of all such universes is rotationally invariant (this is sometimes called ``statistically isotropic" in the CMB community).  
In addition, the two-point correlation function, which describes correlations in the GW power on the sky, is especially simple for the standard Gaussian ensemble; namely, it has the same value {\it independent} of the angular separation between the two different directions~\footnote{The two-point {\em covariance} function $\langle \psi(\hatOm)\psi(\hatOm')\rangle_\psi - \langle \psi(\hatOm)\rangle_\psi \langle \psi(\hatOm')\rangle_\psi$ is identically zero for the standard Gaussian ensemble.}.

To test if our universe has {\em nontrivial} correlated GW power---i.e., that the two-point function has different values depending on the angular separation between two directions on the sky, we construct the following (non-Gaussian) ensemble defined by the following two-stage ensemble averaging process:
 (i) The first stage consists of averaging over Gaussian (sub)ensembles for fixed
anisotropic distributions of GW power on the sky $\psi(\hatOm)$; (ii) the second stage consists 
of averaging over a rotationally invariant ensemble of such $\psi(\hatOm)$'s.
More explicitly, 
given an anisotropic distribution of GW power $\psi(\hatOm)$, we first assume that the Fourier domain strain coefficients obey 
a multivariate Gaussian distribution with zero mean  
\begin{equation}
\langle\tilde{h}_A(f,\hatOm)\rangle=0\,,
\end{equation}
and quadratic expectation values:
\begin{equation}\label{eq:strainSecondMom}
    \langle \tilde{h}_A(f,\hatOm)\,\tilde{h}^*_{A'}(f',\hatOm') \rangle = \psi(\hatOm) H(f)\, \delta_{AA'} \delta(f-f')  \delta^2(\hatOm,\hatOm')\,,
\end{equation}
where $H(f)$ is the real (2-sided, $H(f)=H(-f)$) power spectrum of the GW signal having sky distribution $\psi(\hatOm)$.
Here, $\langle\ \rangle$ without any subscript denotes averaging over the Gaussian ensemble of Fourier coefficients for a fixed $\psi(\hatOm)$.
Stationarity in time and homogeneity in space introduce the delta functions in frequency and angular parameters~\cite{romano2023answers}.  
The fact that the right-hand side (rhs) depends on the polarization components only via $\delta_{AA'}$ means that the polarization components are statistically equivalent and independent of one another.

The nontrivial correlations in the GW power are encoded in the assumptions we make on the statistical distribution of the allowed set of functions $\psi(\hatOm)$.
We will assume (on average) that these functions have no preferred direction on the sky, 
and that the covariance between the GW power in two different directions $\hatOm$, $\hatOm'$ depends only on the angular separation between those two directions
\begin{equation}
\begin{aligned}
\label{eq:def_statsIso_pix}
&\langle \psi(\hatOm)\rangle_\psi=1\,,
\\
&\langle \psi(\hatOm)\,\psi(\hatOm') \rangle_\psi
-\langle \psi(\hatOm)\rangle_\psi\langle \psi(\hatOm')\rangle_\psi
=C(\hatOm\cdot\hatOm')\,.
\end{aligned}
\end{equation}
Here, $\langle\ \rangle_\psi$ denotes the ensemble average over the set of allowed functions $\psi(\hatOm)$. 
Using \eqref{eq:strainSecondMom} and \eqref{eq:def_statsIso_pix}, it immediately follows that
\be
\langle\langle \tilde{h}_A(f,\hatOm)\,\tilde{h}^*_{A'}(f',\hatOm') \rangle\rangle_\psi= H(f)\, \delta_{AA'} \delta(f-f')  \delta^2(\hatOm,\hatOm')\,,
\label{e:hh_U}
\ee
where the left-hand side (lhs) is obtained by first applying the Gaussian ensemble average $\langle\ \rangle$ over the $\tilde h_A(f,\hatOm)$'s and then the ensemble average $\langle\ \rangle_\psi$ over the $\psi(\hatOm)$'s.
Note that the final result is the same as the quadratic expectation value of the $\tilde h_A(f,\hatOm)$'s for the standard Gaussian ensemble.

Since the cosmic variance calculation that we perform will involve at most 4th-order expectation values of the $\tilde h_A(f,\hatOm)$'s, it suffices to know only the first two moments of the $\psi(\hatOm)$'s.
To illustrate this statement, let us use the following shorthand notation
\be
\begin{aligned}
&h_1\equiv \tilde h_{A_1}(f_1,\hatOm_1)\,,\quad
h_2^*\equiv \tilde h_{A_2}^*(f_2,\hatOm_2)\,,\quad
H_1\equiv H(f_1)\,,\quad
\psi_1\equiv \psi(\hatOm_1)\,,\quad
C_{12}\equiv C(\hatOm_1\cdot\hatOm_2)\,,
\\
&\delta_{12}\equiv\delta_{A_1 A_2}\delta(f_1-f_2)\delta^2(\hatOm_1,\hatOm_2)\,,\quad \delta_{13}^*\equiv \delta_{A_1 A_3}\delta(f_1+f_3)\delta^2(\hatOm_1,\hatOm_3)\,, \quad{\rm etc.}
\label{e:shorthand}
\end{aligned}
\ee
Then
\be
\begin{aligned}
\label{e:hhhh_U}
\langle\langle h_1 h_2^* h_3 h_4^*\rangle\rangle_\psi
&=\langle
\langle h_1 h_2^*\rangle\langle h_3 h_4^*\rangle+
\langle h_1 h_3\rangle\langle h_2^* h_4^*\rangle+
\langle h_1 h_4^*\rangle\langle h_2^* h_3\rangle\rangle_\psi
\\
&=
\langle \delta_{12}H_1\psi_1\delta_{34}H_3\psi_3\rangle_\psi+\langle\delta_{13}^*H_1\psi_1\delta_{24}^*H_2\psi_2\rangle_\psi+\langle\delta_{14}H_1\psi_1\delta_{23}H_2\psi_2\rangle_\psi
\\
&=
\delta_{12}\delta_{34}H_1H_3\langle \psi_1\psi_3\rangle_\psi+\delta_{13}^*\delta_{24}^*H_1H_2\langle \psi_1\psi_2\rangle_\psi+\delta_{14}\delta_{23}H_1 H_2\langle \psi_1\psi_2\rangle_\psi
\\
&=\delta_{12}\delta_{34}H_1H_3(C_{13}+1)+
\delta_{13}^*\delta_{24}^*H_1H_2(C_{12}+1)+
\delta_{14}\delta_{23}H_1H_2(C_{12}+1)
\\
&=\langle h_1 h_2^* h_3 h_4^*\rangle_{\rm gauss} + 
\delta_{12}\delta_{34}H_1H_3C_{13}+
\delta_{13}^*\delta_{24}^*H_1H_2 C_{12}+
\delta_{14}\delta_{23}H_1H_2 C_{12}\,.
\end{aligned}
\ee
Note that we first do the $\langle\ \rangle$ Gaussian ensemble average over the four $h$'s for fixed $\psi(\hatOm)$, 
using Isserlis's theorem for zero-mean Gaussian random variables to expand $\langle h_1 h_2^* h_3 h_4^*\rangle$.
We then do the $\langle\ \rangle_\psi$ ensemble average over the $\psi(\hatOm)$'s to obtain the third and fourth equalities using \eqref{eq:def_statsIso_pix}.
The expression $\langle h_1 h_2^* h_3 h_4^*\rangle_{\rm gauss}$
in the final line is the equivalent 4th-order expectation values for the standard Gaussian ensemble.

Furthermore, it is generically useful to decompose the GW power on the sky in terms of spherical harmonics. For example, the expectation values \eqref{eq:def_statsIso_pix} for the $\psi(\hatOm)$'s can be expressed in terms of the spherical harmonic components $\psi_{l m}$ of a given sky map $\psi(\hatOm)$ defined by: 
\begin{equation}\label{eq:SpH_decompose}
    \psi(\hatOm) = \sum_{l=0}^{\infty}\sum_{m=-l}^{l}\,\psi_{l m}\,Y_{l m}(\hatOm)\,,\qquad
    \psi_{l m} =\int {\rm d}^2\hatOm\> \psi(\hatOm)\,Y^*_{l m}(\hatOm)\,,
\end{equation}
where we adopt the normalization of spherical harmonics used by Arfken 2005 (p. 791,~\cite{Arfken}). Since spherical harmonics satisfy the following orthonormality condition and addition theorem, 
\begin{align}
\label{eq:SpH_OrthoNormal_Cond}
\int {\rm d}^2\hatOm\> Y_{l m} (\hatOm)Y^{*}_{l' m'} (\hatOm) &=\delta_{l l'} \delta_{mm'}\,,\\
\label{e:addition_thm}
\sum_{m=-l}^{l} Y_{l m} (\hatOm)Y^{*}_{l m} (\hatOm') &=  \frac{2l +1}{4\pi}P_l(\hatOm\cdot\hatOm')\,,
\end{align}
it follows that
\begin{equation}
\begin{aligned}
    \label{eq:SpH_Stats}
    \langle \psi_{l m} \rangle_\psi &=\sqrt{4\pi}\, \delta_{l 0} \delta_{m0}\,,\\
    \langle \psi_{l m} \psi^*_{l' m'}\rangle_\psi-\langle \psi_{l m} \rangle_\psi\langle \psi^*_{l' m'} \rangle_\psi &= C_l \delta_{l l'}\delta_{mm'}\,,
\end{aligned}
\end{equation}
where $P_l(x)$ in \eqref{e:addition_thm} is the $l$th order Legendre polynomial. 
The angular power spectrum $C_l$ is defined in terms of $C(\hatOm\cdot\hatOm')$ via 
\begin{equation}\label{eq:Cpix2CSph}
   C(\hatOm\cdot\hatOm') = \sum_{l=0}^\infty\,\frac{2l+1}{4\pi}  C_l P_l (\hatOm\cdot\hatOm')\qquad{\rm or}\qquad C(\hatOm\cdot\hatOm')=\sum_{l,m} C_l Y_{l m}(\hatOm)Y_{l m}^*(\hatOm')\,,
   \end{equation}
where we have introduced the notation $\sum_{l,m}=\sum_{l=0}^{\infty} \sum_{m=-l}^{l}$. A {\em white} angular power spectrum has $C_l={\rm const}$ for all $l$ values up to some $l_{\rm max}$.
A {\em scale-invariant} angular power spectrum, which has equal variance per logarithmic spacing in $l$, has $l(l+1)C_l={\rm const}$.
Finally, we note that $\langle \psi(\hatOm)\psi(\hatOm')\rangle_\psi=C(\hatOm\cdot\hatOm')+1$ has Legendre polynomial coefficient $C_l+4\pi\delta_{l 0}$.

Since the spectrum $H(f)$ and the angular power spectrum $C_l$ for $l=0,1,\cdots$ are sufficient to describe rotationally invariant correlations in GW power, we are now ready to derive the expected behavior of the redshift correlation curve for PTA experiments. 

\section{Cosmic variance calculation}\label{sec:CV_Position}

Here, we are interested in accounting for the effect of nontrivial correlations in GW power on the mean and variance of the pulsar pair correlation
\begin{equation}
\begin{aligned}
    \label{eq:def_rho}
    \rho_{12} &= \overline{z_1(t)\,z_2(t)} \\
    &=\sum_{A,A'}\int \d f\,\int \d f'\,\int_{S^2}\d^2\hatOm\,\int_{S^2}\d^2\hatOm' \,\tilde{h}_A(f,\hatOm)\tilde{h}^*_{A'}(f',\hatOm')\,R_1^A(f,\hatOm)R_2^{A'*}(f',\hatOm')\,\textrm{sinc}[\pi T (f-f')]\,,
\end{aligned}
\end{equation}
where we used \eqref{eq:redshift} for the redshifts, with $R^A_a(f,\hatOm)\equiv R^A(f,\hatOm, D_a\,\hatp_a)$ for $a=1,2$ labeling the two pulsars.
We will assume that we have access to an infinite number of noise-free pulsar redshift measurements, which we can bin by the separation angle between pairs of pulsars and then average the correlations together.  This averaging over pulsar directions, denoted 
$\langle\ \rangle_{12\in\gamma}$, keeps the separation angle $\gamma\equiv \cos^{-1}(\hatp_1\cdot\hatp_2)$ between the two pulsars fixed. 
This averaging is appropriate when constructing estimates of the GWB that are 
rotationally invariant, i.e., that depend only on the angular separation between two points on the sky.  
If, instead, we wanted to map the anisotropy $\psi(\hatOm)$ of the GWB in our particular realization of 
the Universe, then we should not average the correlation measurements having the same angular separation.

As mentioned earlier, pulsar averaging removes the pulsar variance, leaving only the cosmic variance.
It also removes any dependence of the correlated response on the distance to the pulsars, provided the correlation length of the GW background is much shorter than the Earth-pulsar and pulsar-pulsar distances~\cite{Allen23}.
For this case, $R^A_a(f,\hatOm)$ can be replaced by $F^A_a(\hatOm)\equiv F^A(\hatOm, \hatp_a)$, leading to
\begin{equation}\label{eq:PA_Gamma}
    \Gamma (\gamma)\equiv \langle \rho_{12} \rangle_{12\in\gamma} = \sum_{A,A'}\int \d f\,\d f'\int \d^2\hatOm\,\d^2\hatOm' \,\tilde{h}_A(f,\hatOm)\tilde{h}^*_{A'}(f',\hatOm')\,\mu_{AA'}(\gamma,\hatOm,\hatOm')\,\textrm{sinc}[\pi T (f-f')] \,,
\end{equation}
where $\mu_{AA'}(\gamma,\hatOm,\hatOm')$ is the Hellings and Downs two-point function~\cite{Allen23}:
\begin{equation}
\label{e:mu_AA'}
\mu_{AA'}(\gamma,\hatOm,\hatOm')\equiv\langle F_1^A(\hatOm)F_2^{A'}(\hatOm')\rangle_{12\in\gamma}\,.
\ee
In terms of the pulsar-averaged correlation $\Gamma$, the cosmic variance is defined as
\be
\label{eq:def_CV}
    \sigma_{\rm cosmic}^2 = \langle\langle  \Gamma^2 \rangle\rangle_\psi - \langle\langle  \Gamma \rangle\rangle_\psi^2\,,
\ee
where the averaging is taken over the two-stage ensemble $\langle\langle\ \rangle\rangle_\psi$.

\subsection{Earth-term-only response function and HD two-point function in harmonic space}
\label{s:Flm}

The Earth-term-only response functions $F^A(\hatOm, \hatp)$ which appears in \eqref{e:mu_AA'}, can be written in harmonic space with respect to pulsar directions on the sky
\be
\label{e:F_lm_def}
F^A(\hatOm, \hatp) 
=\sum_{l, m}
F^A_{lm}(\hatOm) Y_{l m}(\hatp)
\quad\Leftrightarrow\quad
F^A_{lm}(\hatOm)=
\int {\rm d}^2\hatp\>
F^A(\hatOm, \hatp)\, Y_{l m}^*(\hatp)\,.
\ee
Note that $F_{l m}^A(\hatOm)$ is complex, satisfying
$F_{l m}^{A*}(\hatOm) = (-1)^m \,F_{l, -m}^A(\hatOm)$
as a consequence of $Y_{l m}^*(\hatOm) = (-1)^{m}\,Y_{l,-m}(\hatOm)$.
As shown in \citet{BernardoPhenomeanology}, the integral in \eqref{e:F_lm_def} can be evaluated as
\begin{equation}\label{eq:RlmFinal_infDistance}
    F_{l m}^{A=+,\times}(\hatOm) = -2\pi i (-i)^{2l-1}_{2l}\,\sqrt{\frac{(l-2)!}{(l+2)!}}\, \left[{}_{-2}Y^*_{l m }(\hatOm)\textrm{e}^{-i2\alpha(\hatOm)}\pm{}_{2}Y^*_{l m }(\hatOm)\textrm{e}^{i2\alpha (\hatOm)}\right]\quad {\rm for } \,l\geq2\,.
\end{equation}
 We note that due to a difference in our definition for the metric perturbations, an extra term of $(-i)^{2l-1}_{2l}$ appears in the above equation. (The notation $(-i)^{2l-1}_{2l}$ mean raise $-i$ to the power of $2l-1$ for $A=+$ and $2l$ for $A=\times$.) Here, $\alpha (\hatOm)$ is the polarization angle for the GW propagating in direction $\hatOm$.  It turns out to be a nuisance parameter for us, as our observables are independent of this parameter.
 
In terms of the $F_{l m}^A(\hatOm)$'s, the Hellings and Downs two-point function can be written as
\be
\mu_{AA'}(\gamma,\hatOm,\hatOm')
=\frac{1}{4\pi}\sum_{l,m}
F_{l m}^A(\hatOm)F_{l m}^{A'*}(\hatOm')\,P_l(\cos\gamma)\,,
\label{e:2pt-harmonic}
\ee
where we used \cite{SkyAverageSpH,Bernardo_2022}
\be
\langle Y_{l m}(\hatp_1)Y_{l' m'}^*(\hatp_2) \rangle_{12\in\gamma}
=\delta_{ll'}\delta_{mm'}\frac{P_l(\cos\gamma)}{4\pi}\,,
\ee
for the pulsar-averaged product of two spherical harmonics. The Hellings and Downs curve $\mu_{\rm u}(\gamma)$ can also be written in terms of the $F_{l m}^A(\hatOm)$'s using~(\ref{e:2pt-harmonic}) and the first line of \eqref{e:sumFlms} as
\be
\begin{aligned}
\mu_{\rm u}(\gamma) \equiv \sum_A \mu_{AA}(\gamma,\hatOm,\hatOm)
&=\frac{1}{4\pi}\sum_{l, m} \sum_A F_{l m}^A(\hatOm)F_{l m}^{A*}(\hatOm)\,P_l(\cos\gamma)\\
&= 2\pi\sum_{l}\frac{(l-2)!}{(l+2)!}\sum_{s=-2,2} \sum_m{}_{s}Y^*_{l m}(\hatOm)\,{}_{s}Y_{l m}(\hatOm)\,P_l(\cos\gamma)\\
&=2\pi \sum_{l} \frac{(l-2)!}{(l+2)!}\sum_{s=-2,2}\frac{2l+1}{4\pi}\,P_l(\cos\gamma)\\
&=\sum_{l} a_l P_l(\cos\gamma)
\,,
\end{aligned}
\label{e:HD-curve}
\ee
where 
\be
a_l \equiv \frac{2l +1}{(l+2)(l+1)l(l-1)}\,.
\label{e:a_ell}
\ee
This Legendre polynomial expansion can be derived using expressions from Appendix~\ref{appendix:RlmRelations}. The fourth line is obtained by using the addition theorem for spin-weighted spherical harmonics [see (2.61) p. 56,~\cite{Torres2003}]. These $a_l$ coefficients appear in multiple expressions that follow. We note that the HD curve $\mu_{\rm u}(\gamma)$ is independent of the direction of the source in the sky~\cite{Allen23}.

\subsection{First moment of the pulsar-averaged correlation}

Since the quadratic expectation values of the $\tilde h_A(f,\hatOm)$'s for ensembles having nonzero angular power spectra are the same as for the standard Gaussian ensemble \eqref{e:hh_U}, it is also true for the mean of the pulsar-averaged correlation:
\be
\begin{aligned}
\mu_{\rm cosmic}&=\langle \langle\Gamma \rangle\rangle_\psi =  \sum_{A_1,A_2}\int \d f_1\,\d f_2\int \d^2\hatOm_1\,\d^2\hatOm_2 \,\textrm{sinc}\left[\pi T (f_1-f_2)\right]\,\mu_{A_1 A_2}(\gamma,\hatOm_1,\hatOm_2)\,\langle\tilde{h}_{A_1}(f_1,\hatOm_1)\tilde{h}^*_{A_2}(f_2,\hatOm_2)\rangle_\psi
\\
&=  \sum_{A_1,A_2}\int \d f_1\,\d f_2\int \d^2\hatOm_1\,\d^2\hatOm_2 \,\textrm{sinc}\left[\pi T (f_1-f_2)\right]\,\mu_{A_1 A_2}(\gamma,\hatOm_1,\hatOm_2)\,H(f_1)\delta_{A_1A_2}\delta(f_1-f_2)\delta^2(\hatOm_1,\hatOm_2)
\\
&=
\sum_{A_1}\int \d f_1\int \d^2\hatOm_1\,\mu_{A_1 A_1}(\gamma,\hatOm,\hatOm)\,H(f_1)
\\
&= h^2\,\mu_u(\gamma)\quad(=\langle\Gamma\rangle_{\rm gauss})\,,
\end{aligned}
\label{e:meanHD}
\ee
where
$h^2 \equiv 4 \pi \int {\rm d}f H(f)$. The last equality is obtained by using~(\ref{e:HD-curve}).

\subsection{Second moment and variance of the pulsar-averaged correlation}

Next, we obtain an expression for the second moment of $\Gamma$
\begin{equation}
    \begin{aligned}
\label{eq:PA_Gamma_squared}
     \langle\langle\Gamma^2 \rangle\rangle_\psi = & \sum_{A_1,A_2,A_3,A_4}\int \d f_1\,\d f_2\,\d f_3\,\d f_4\,
     \int \d^2\hatOm_1\,\d^2\hatOm_2 \,\d^2\hatOm_3\,\d^2\hatOm_4
     \\
     &\hspace{0.25in}\times \textrm{sinc}\left[\pi T (f_1-f_2)\right]\,\textrm{sinc}\left[\pi T (f_3-f_4)\right]\mu_{A_1A_2}(\gamma,\hatOm_1,\hatOm_2)\,\mu_{A_3A_4}(\gamma,\hatOm_3,\hatOm_4)\\
     &\hspace{0.25in}\times\langle\langle\tilde{h}_{A_1}(f_1,\hatOm_1)\tilde{h}^*_{A_2}(f_2,\hatOm_2)\tilde{h}_{A_3}(f_3,\hatOm_3)\tilde{h}^*_{A_4}(f_4,\hatOm_4)\rangle\rangle_\psi\,.
\end{aligned}
\end{equation}
To facilitate the calculation, we will use the shorthand notation introduced in \eqref{e:shorthand} supplemented with
\be
\begin{aligned}
&\int_{1,2,3,4}\equiv\sum_{A_1,A_2,A_3,A_4}\int {\rm d}f_1\,{\rm d}f_2\,
{\rm d}f_3\,{\rm d}f_4
\int{\rm d}^2\hatOm_1\, {\rm d}^2\hatOm_2\,{\rm d}^2\hatOm_3\, {\rm d}^2\hatOm_4\,,
\end{aligned}
\ee
as well as ${\rm sinc}_{12}$ and $\mu_{12}$, etc., for the sinc and HD two-point functions, respectively.
In this notation
\be
\begin{aligned}
\langle\langle\Gamma^2 \rangle\rangle_\psi &=\int_{1,2,3,4}\>
{\rm sinc}_{12}\,
{\rm sinc}_{34}\,\mu_{12}\,\mu_{34}\,\langle\langle h_1 h_2^* h_3 h_4^*\rangle\rangle_\psi\\ 
&=
\langle\Gamma^2\rangle_{\rm gauss}+
\int_{1,3}\>
\mu_{11}\,\mu_{33}\,H_1 H_3\,C_{13}
+2\int_{1,2}\>
{\rm sinc}^2_{12}\,
\mu^2_{12}\,H_1\,H_2\,C_{12}\,,
\label{e:G2_psi}
\end{aligned}
\ee
where we used \eqref{e:hhhh_U} and symmetry properties of the sinc function and HD two-point function to get the second line
of \eqref{e:G2_psi}
 (the first two terms on the second line of \eqref{e:G2_psi} correspond to the
first two terms of \eqref{e:hhhh_U} with the delta functions enforcing that indices 2 and 4 
are replaced by 1 and 3; while the final term on the second line of \eqref{e:G2_psi} comes from 
combining the last two terms of \eqref{e:hhhh_U} with the delta functions enforcing that 
indices 3 and 4 are replaced by 1 and 2 and 2 and 1, respectively.)
Since $\langle\langle \Gamma\rangle\rangle_\psi=\langle\Gamma\rangle_{\rm gauss}$, it follows that 
\be
\sigma^2_{\rm cosmic}=
\sigma^2_{\rm cosmic,\, gauss}
+\int_{1,3}\>
\mu_{11}\,\mu_{33}\,H_1 H_3\,C_{13}
+2\int_{1,2}\>
{\rm sinc}^2_{12}\,
\mu^2_{12}\,H_1\,H_2\,C_{12}\,,
\label{e:4.15}
\ee
where 
\be
\sigma^2_{\rm cosmic,\, gauss} \equiv \langle\Gamma^2\rangle_{\rm gauss} - \langle\Gamma\rangle_{\rm gauss}^2 = 2{\cal h}^4\, \widetilde{\mu^2}(\gamma)\,
\ee
is the cosmic variance for the standard Gaussian ensemble.\\
Here,
\begin{equation}
\mathcal{h}^4 = (4 \pi)^2\int \d f_1\int \d f_2 \,\textrm{sinc}^2\left[\pi T (f_1-f_2)\right]\,H(f_1) H(f_2)\,.
\label{e:htilde4}
\end{equation}
We now evaluate the two ``integrals" on the rhs of the \eqref{e:4.15}. The first is equal to 
\be
I_1 \equiv
\int_{1,3}\>
\mu_{11}\,\mu_{33}\,H_1 H_3\,C_{13}
= \mu^2_{\rm u}(\gamma) \,h^4\, \int\frac{{\rm d}^2\hatOm_1}{4\pi}
\int\frac{{\rm d}^2\hatOm_3}{4\pi}
C(\hatOm_1\cdot\hatOm_3)\,.
\ee
Recall that $\mu_{\rm u}(\gamma)=\sum_A\mu_{AA}(\gamma,\hatOm,\hatOm)$ and $h^2 \equiv 4\pi\int {\rm d}f\, H(f)$.  The double integral over sky directions can be simply evaluated since $C(\hatOm_1\cdot\hatOm_2)$ depends only on the dot product of $\hatOm_1$ and $\hatOm_2$:
\begin{equation}
\int\frac{\d^2\hatOm_1}{4\pi}\int\frac{\d^2\hatOm_3}{4\pi}\, C(\hatOm_1\cdot\hatOm_3)
= \frac{1}{2}\int_{-1}^1 {\rm d}x \>C(x)
=\frac{1}{2}\int_{-1}^1 {\rm d}x\> \sum_{l=0}^\infty\frac{2l+1}{4\pi}\,C_l P_l(x) = \frac{1}{4\pi}\,C_0\,.
\ee
Thus,
$I_1 = (h^4/4\pi)\,C_0 \,\mu_{\rm u}^2(\gamma)$.
The second integral is equal to 
\be
\begin{aligned}
I_2 &\equiv
2\int_{1,2}\>
{\rm sinc}^2_{12}\,
\mu^2_{12}\,H_1 H_2\,C_{12}
\\
&= 2\,(4\pi)^2
\int {\rm d}f_1\int {\rm d}f_2\,{\rm sinc}^2[\pi T(f_1-f_2)]\,H(f_1) H(f_2) \int\frac{{\rm d}^2\hatOm_1}{4\pi}
\int\frac{{\rm d}^2\hatOm_2}{4\pi}\,
\sum_{A_1,A_2}\mu^2_{A_1 A_2}(\gamma, \hatOm_1,\hatOm_2)
C(\hatOm_1\cdot\hatOm_2)\,.
\end{aligned}\ee
To evaluate the integral over sky directions, we express the HD two-point function in harmonic space [see \eqref{e:2pt-harmonic}] and expand $C(\hatOm_1\cdot\hatOm_2)$ in terms of spherical harmonics [see \eqref{eq:Cpix2CSph}].
This yields
\be
\begin{aligned}
&\int\frac{{\rm d}^2\hatOm_1}{4\pi}
\int\frac{{\rm d}^2\hatOm_2}{4\pi}\,
\sum_{A_1,A_2}\mu^2_{A_1 A_2}(\gamma, \hatOm_1,\hatOm_2)
C(\hatOm_1\cdot\hatOm_2)\\
&\hspace{0.25in}=\sum_{L,M} C_L \sum_{l,m}\sum_{l',m'}\frac{P_l(\cos\gamma)}{4\pi} \frac{P_{l'}(\cos\gamma)}{4\pi} \left| \frac{1}{4\pi}\int {\rm d}^2\hatOm\, Y_{LM}(\hatOm)\sum_A F_{l m}^A(\hatOm) F_{l' m'}^{A*}(\hatOm) \right|^2
\\
&\hspace{0.25in}=
\sum_{L,M} C_L \sum_{l,m}\sum_{l',m'}\frac{P_l(\cos\gamma)}{4\pi} \frac{P_{l'}(\cos\gamma)}{4\pi}
\pi\,a_l a_{l'}(2L+1)
\begin{pmatrix}
l & l' & L\\
-m & m' & M
\end{pmatrix}^2
\big[ 1 + (-1)^{l+l'+L}\big]^2\begin{pmatrix}
l & l' & L\\
2 & -2 & 0
\end{pmatrix}^2
\\
&\hspace{0.25in}=\frac{1}{8\pi}\sum_L (2L+1)C_L
\sum_{l,l'}
a_l P_l(\cos\gamma)a_{l'}P_{l'}(\cos\gamma)
\big[ 1 + (-1)^{l+l'+L}\big]\begin{pmatrix}
l & l' & L\\
2 & -2 & 0
\end{pmatrix}^2\,,
\end{aligned}
\ee
where we used
\eqref{eq:YlmRlm2Int_1} and \eqref{e:a_ell} to get the second equality, 
and
\begin{equation}
   \sum_{m, m'} (2L+1)\begin{pmatrix}
l & l' & L\\
-m & m' & M
\end{pmatrix}^2
=1\,,
\end{equation}
which is a consequence of \eqref{eq:Wigner-3j_Orthogonal}, to get the third equality.
Combining the above results, the cosmic variance takes its final form:
\be
\begin{aligned}
\sigma^2_{\rm cosmic} &= 2{\cal h}^4\,\widetilde{\mu^2}(\gamma)+ \frac{h^4}{4\pi}\,C_0\,\mu_{\rm u}^2(\gamma)
\\
&\hspace{0.5in}
+\frac{{\cal h}^4}{4\pi}
\sum_L (2L+1)C_L
\sum_{l,l'}
a_l P_l(\cos\gamma)a_{l'}P_{l'}(\cos\gamma)
\left[ 1 + (-1)^{l+l'+L}\right]\begin{pmatrix}
l & l' & L\\
2 & -2 & 0
\end{pmatrix}^2\,.
\label{e:CV_Final}
\end{aligned}
\ee
The first term is the cosmic variance for the standard Gaussian ensemble,  while the other two terms, proportional to $C_0$ and the summation over the $C_l$'s, are the additional contributions to the cosmic variance due to nonzero values for the angular power spectrum $C_l$.
If desired, the cosmic variance can be expressed completely in terms of harmonic coefficients using \eqref{e:hd-harmonic} and \eqref{e:tildemu2-harmonic} for $\mu_{\rm u}(\gamma)$ and $\widetilde{\mu^2}(\gamma)$.

Plots of the expected cosmic variance for the different ensembles are shown in Fig.~\ref{fig:CVStatsIso}. For making these plots, we assumed: (i) $h^2=1$ and $\mathcal{h}^4=0.5$. (ii) For a white angular power spectrum, $C_L=C_0$ for $L\le L_{\rm max}=8$. (iii) For a scale-invariant angular power spectrum, $L (L+1) C_L = \text{const} =C_0$ for $1\leq L \leq L_{\rm max}=8$. We used $l$ values up to $l_{\rm max}=50$ to construct the detector response. We note that for large values of the angular power spectra (i.e., $C_0=1$), the expected cosmic variance differs visually from that for the standard Gaussian ensemble, while for more realistic values (i.e., $C_0=10^{-3}$), there is no visual distinction.
\begin{figure}[t]
    \centering
\includegraphics[width=0.45\textwidth]{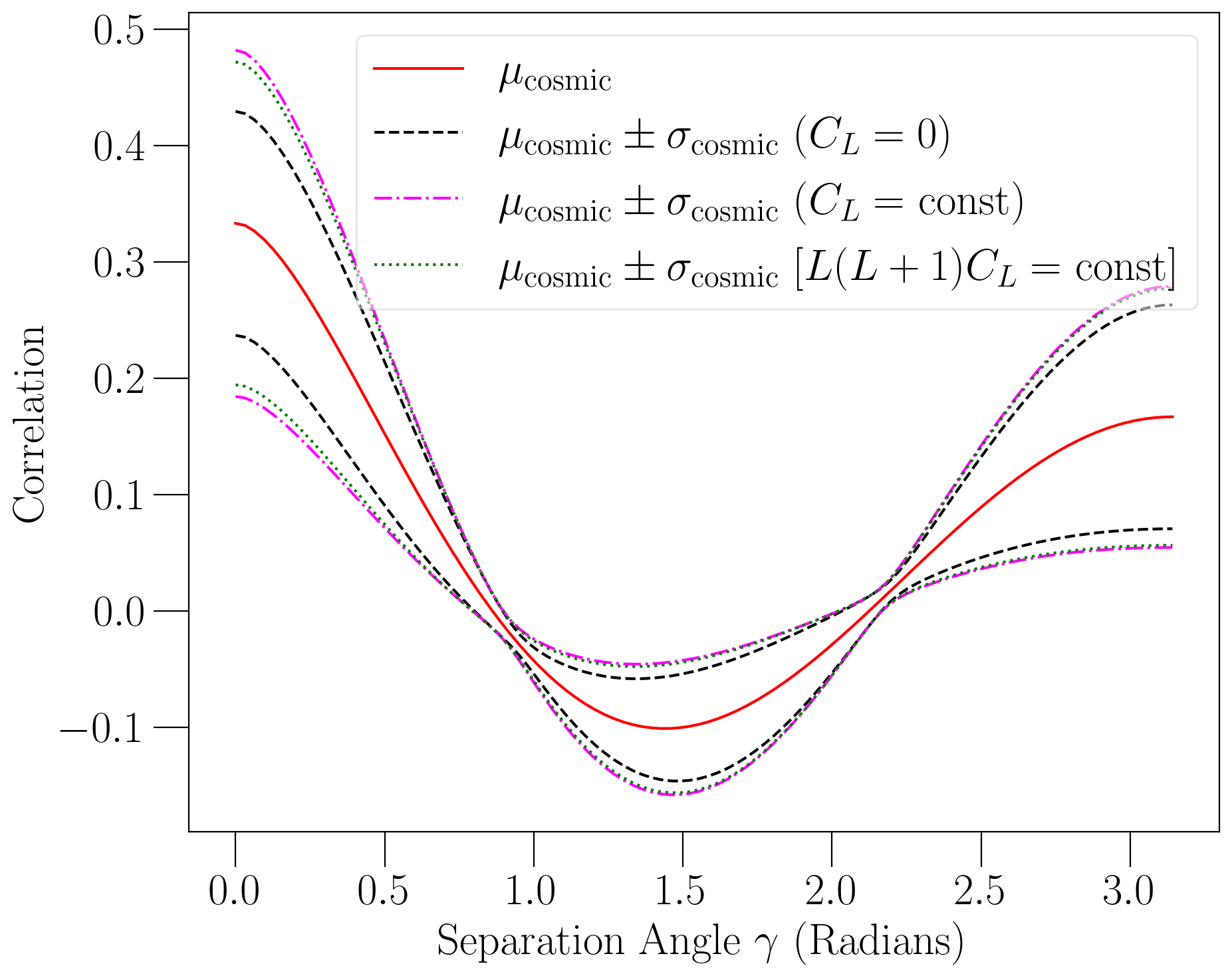}~~~\includegraphics[width=0.45\textwidth]{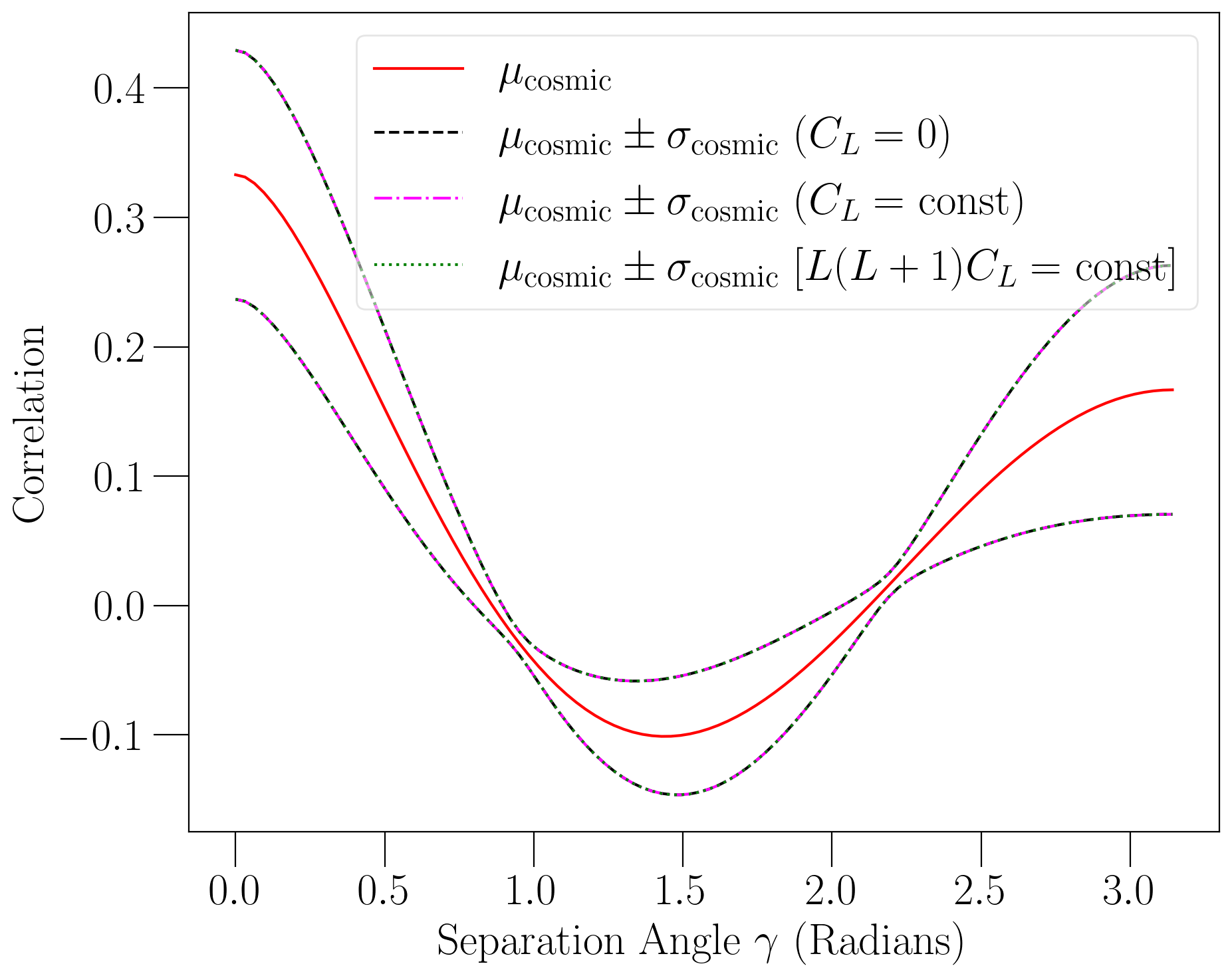}
    \caption{Expected correlation and cosmic variance for the standard Gaussian ensemble and ensembles described by nonzero angular power spectra. The mean~(\ref{e:meanHD}) of the correlation curve (HD curve) is shown by the red solid line setting $h^2=1$. The dashed black, dashed-dotted magenta, and dotted green lines represent the 1-sigma expected deviation~(\ref{e:CV_Final}) from the HD curve for the standard Gaussian ensemble and for ensembles having white ($C_L={\rm const}$) and scale invariant ($L(L+1)C_L={\rm const}$) angular power spectra, respectively. 
We have set $\mathcal{h}^4/h^4=1/2$, and used detector modes up to $l_{\rm max}=50$ and signal modes up to $L_{\rm max}=8$. 
 (The choice $\mathcal{h}^4/h^4=1/2$ is appropriate for GW sources all radiating in a single frequency bin
with that frequency commensurate with the inverse of the total observation time.)
The scale of the angular power spectrum is set by $C_0=1$ and $C_0=10^{-3}$ in the left and right panels, respectively.}
    \label{fig:CVStatsIso}
\end{figure}  

\section{Discussion}\label{sec:Conclusions}

The aim of this article was to assess how nontrivial correlations in the GW power on the sky affect the recovery of the HD correlation curve, assuming an infinite number of noise-free pulsars.
We utilized a harmonic space decomposition of the detector response to facilitate the analytical calculation of the first two moments of the pulsar-averaged correlation curve, i.e., its mean and (cosmic) variance. We constructed a simplistic two-stage ensemble process for characterizing the correlations in the GW power, specified by a monopole $H(f)$ and angular power spectrum $C_l$. We found that the mean correlation (\ref{e:meanHD}) carries no information about the correlations, while the variance~(\ref{e:CV_Final}) has additional contributions if the angular power spectrum $C_l$ describing the correlations in GW power is nonzero. For the Gaussian ensemble, the variance arises solely due to interference of the GW sources. 
Plots of the expected cosmic variance for the different ensembles described in this paper are given in Fig.~\ref{fig:CVStatsIso}.

To further quantify the effect of a nonzero angular power spectrum, we construct an estimator of the deviation in the cosmic variance for both types of ensembles discussed in this paper via
\be
\Delta_{\chi^2} = \bigg|1-\frac{1}{N}\sum_{i=1}^{N} \frac{\sigma^2_{{\rm cosmic},i}}{\sigma^2_{{\rm cosmic}, {\rm gauss},i}}\bigg|\,,
\label{e:Delta}
\ee
which is motivated by a $\chi^2$ statistic. Recall that a $\chi^2$ statistic is often employed to compare the observation of a measured curve with a theoretical curve, which in this case is the HD correlation curve. 
If the separation angles for pulsar pairs are divided into $N$ bins, we can compare the squared deviation of the measured and expected correlations $x_i$ and $\mu_{{\rm cosmic},i}$ to the cosmic variance for both ensembles
\be
\chi^2=\sum_{i=1}^{N}\,\frac{(x_i-\mu_{{\rm cosmic},i})^2}{\sigma^2_{{\rm cosmic},i}}\,,
\qquad
\chi_{\rm gauss}^2=\sum_{i=1}^{N}\,\frac{(x_i-\mu_{{\rm cosmic},i})^2}{\sigma^2_{{\rm cosmic},{\rm gauss},i}}\,.
\ee
The expected fractional difference in these two $\chi^2$ statistics is then
\be
\Delta_{\chi^2}=\frac{|\langle\langle\chi^2-\chi_{\rm gauss}^2\rangle\rangle_\psi|}{\langle\langle \chi^2\rangle\rangle_\psi}\,,
\ee
which can be written in terms of cosmic variance~(\ref{e:Delta}). We plot this difference as a function of the signal parameters in Fig.~\ref{fig:DeltaWhite} using $N=100$ bins. The difference lies in the range $\Delta_{\chi^2} \sim 5\times 10^{-4}-0.62$.
\begin{figure}[t]
    \centering
    \includegraphics[width=\textwidth]{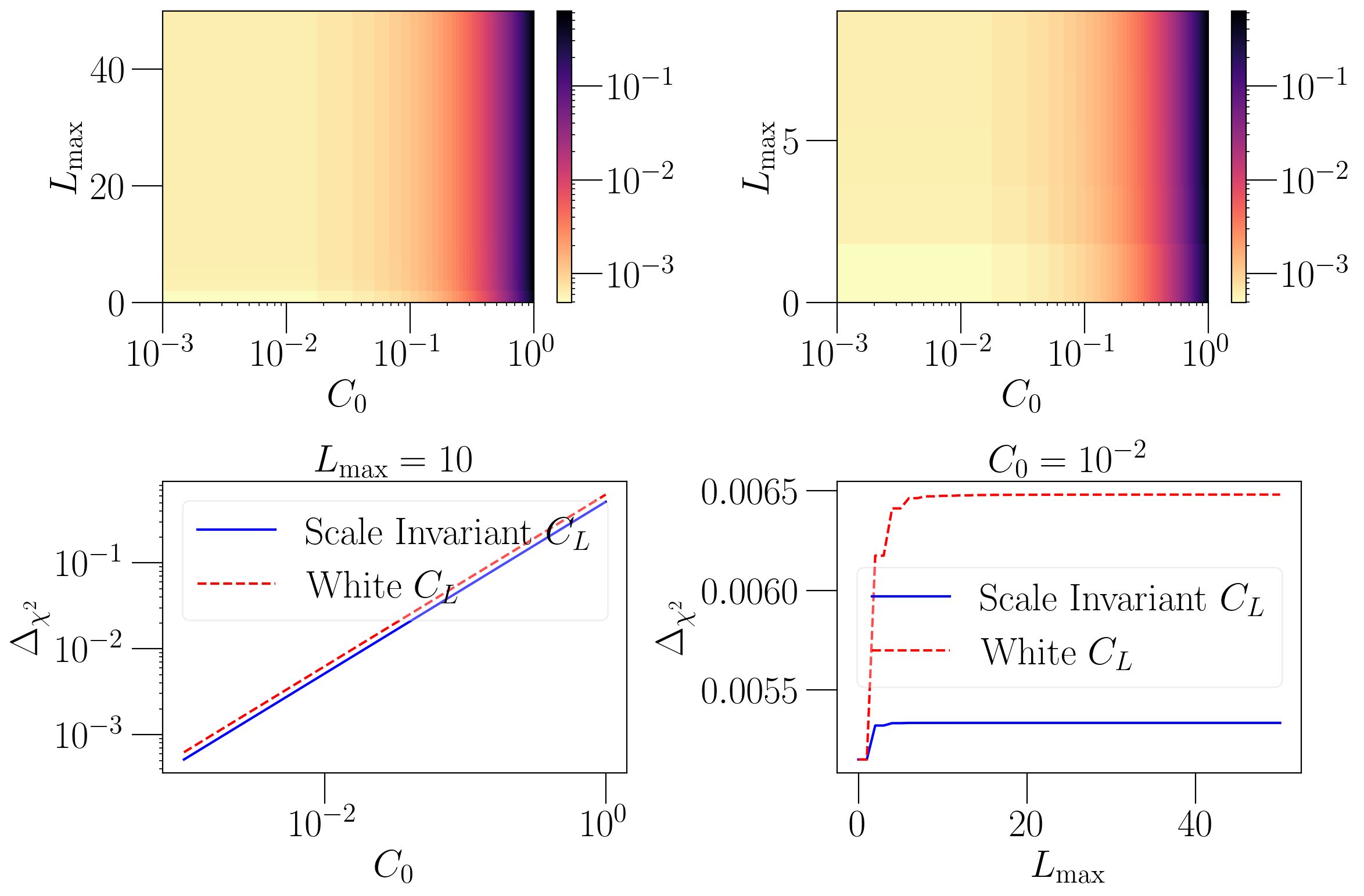}
    \caption{Top left panel: Expected fractional difference $\Delta_{\chi^2}$ in the $\chi^2$ statistic for correlations in GW power defined by a white angular power spectrum with a range of values for $C_0$ and $L_{\rm max}$. Top right panel: Zoom-in of the top-left panel plot for $L_{\rm max}$ between 0 and 10. Bottom left panel: Cross-section of the plot in the top left/top right panel for fixed $L_{\rm max}=10$. Bottom right panel: Cross section of the plot in the top-left panel for fixed $C_0=10^{-2}$.}
    \label{fig:DeltaWhite}
\end{figure}  

Although the effect of a nonzero angular power spectrum on cosmic variance is probably negligible for realistic values of the $C_l$'s, the actual scale of the $C_l$'s is uncertain. Predictions for the angular power spectrum associated with large-scale galaxy clustering are underway~\cite{grimm2024impact}. The expression we provide for the cosmic variance \eqref{e:CV_Final} for an ensemble of universes having rotationally invariant correlations in GW power is general. This expression can be used to estimate the cosmic variance given a model for the angular power spectrum. However, issues like shot noise due to a finite number of galaxies and/or a finite number of GW sources~\cite{PhysRevD.100.063508,satopolito2023exploring,1994ApJ...426...23F}, as well as noise sources in PTA observations might be hurdles in uncovering these angular correlations. With future high-precision PTA experiments coming online, it is crucial to account for all fundamental contributions to the variance while evaluating the timing residual correlations, e.g., the presence of nonzero angular power spectra, anisotropies, polarized GWs, non-Einstein gravity modes, etc. 

NOTE: As this paper was being completed, we became aware of similar work by Bruce Allen~\cite{Allen:2024}, where (among other related topics) he also calculates the cosmic variance for an ensemble of universes having nonzero angular power spectra. The paper of Allen is more extensive in its scope and presentation, e.g., he carefully lays out the ensemble averaging process and derives expressions for the full covariance matrix for a finite number of pulsar pairs, etc.~\cite{Allen:2024}. The final expressions are the same for the subset of topics where our analysis and the analysis by Allen overlap.

\begin{acknowledgments}
D. A. acknowledges financial support from the Actions de Recherche Concertées (ARC) and Le Fonds spécial pour la recherche (FSR) of the Féderation Wallonie-Bruxelles. J. D. R. acknowledges 
financial support from the NSF Physics Frontier Center Award
PFC-2020265 and start-up funds from the University of Texas Rio Grande Valley. He also acknowledges discussions with Bruce Allen and Neil Cornish on this topic at the very early stages of this analysis, and then with Bruce Allen again after the analysis was completed. Bruce pointed out that the full ensemble for nontrivial correlations in GW power is not Gaussian but rather an ensemble of Gaussian subensembles. D. A. and J. D. R. appreciate the efforts of Bruce Allen to help unify the notation between their paper and his~\cite{Allen:2024}, which should make it easier for readers to compare common results.
\end{acknowledgments}


\appendix

\section{Properties of spherical harmonics and Wigner-3j symbols}\label{appendix:Wigner3j}

The triple integral for spin-weighted spherical harmonics, when $s_1+s_2+s_3=0$, is given as
\begin{equation}\label{eq:SpHTripleInt}
    \int \,\d^2\hat{p}\,{}_{s_1}Y_{l_1,m_1}(\hat{p})\,{}_{s_2}Y_{l_2,m_2}(\hat{p})\,{}_{s_3}Y_{l_3,m_3}(\hat{p})=\sqrt{\frac{(2l_1+1)(2l_2+1)(2l_3+1)}{4\pi}}\begin{pmatrix}
l_1 & l_2 & l_3\\
-s_1 & -s_2 & -s_3
\end{pmatrix}
\begin{pmatrix}
l_1 & l_2 & l_3\\
m_1 & m_2 & m_3
\end{pmatrix}\,,
\end{equation}
where $\begin{pmatrix}
a & b & c\\
d & e & f
\end{pmatrix}$ is a Wigner-3j symbol. The Wigner-3j symbol $\begin{pmatrix}
l_1 & l_2 & l_3\\
m_1 & m_2 & m_3
\end{pmatrix}$ vanishes unless $|l_1-l_2|<l_3 <l_1+l_2 $ and $m_1+m_2+m_3=0$. If $m_1=m_2=m_3=0$, then $l_1+l_2+l_3$ must be an even integer. The Wigner-3j symbol satisfies a ``reflection" symmetry
\begin{equation}\label{eq:Wigner3j_reflection}
 \begin{pmatrix}
l_1 & l_2 & l_3\\
m_1 & m_2 & m_3
\end{pmatrix} = (-1)^{l_1+l_2+l_3} \,\begin{pmatrix}
l_1 & l_2 & l_3\\
-m_1 & -m_2 & -m_3
\end{pmatrix}  \,. 
\end{equation}
Some other properties that we will use are
\begin{equation}\label{eq:Wigner-3j_Samelm}
    \begin{pmatrix}
l & l' & 0\\
m & -m' & 0
\end{pmatrix} = \frac{(-1)^{l+m}}{\sqrt{2l+1}}\,\delta_{ll'}\delta_{mm'}\,,
\end{equation}
\begin{equation}\label{eq:Wigner-3j_Sumlm}
   \sum_{m} (-1)^{l-m}\begin{pmatrix}
l & l & L\\
m & -m & 0
\end{pmatrix}=\sqrt{2l+1}\,\delta_{L0}\,,
\end{equation}
\begin{equation}\label{eq:Wigner-3j_Orthogonal}
   \sum_{m_1,m_2} (2L+1)\begin{pmatrix}
l_1 & l_2 & L\\
m_1 & m_2 & M
\end{pmatrix}\,\begin{pmatrix}
l_1 & l_2 & L'\\
m_1 & m_2 & M'
\end{pmatrix}=\delta_{LL'}\delta_{MM'},.
\end{equation}

\section{Useful expressions involving the Earth-term-only response function $F^A_{l m}(\hatOm)$}\label{appendix:RlmRelations}

Using \eqref{eq:RlmFinal_infDistance}, it follows that
\be
\begin{aligned}
   \sum_{A=+,\times} F_{l m}^A(\hatOm)\,F_{l' m'}^{A*}(\hatOm) &= 8\pi^2(-1)^{l+l'}\sqrt{\frac{(l-2)!(l'-2)!}{(l+2)!(l'+2)!}}\sum_{s=-2,2}\,{}_{s}Y^*_{l m}(\hatOm)\,{}_{s}Y_{l' m'}(\hatOm)\\
   &=8\pi^2(-1)^{l+l'+m}\sqrt{\frac{(l-2)!(l'-2)!}{(l+2)!(l'+2)!}}\sum_{s=-2,2} \,{}_{-s}Y_{l, -m}(\hatOm)\,{}_{s}Y_{l' m'}(\hatOm)\,.
\end{aligned}
\label{e:sumFlms}
\end{equation}
Using this last equation, the identity for the integration of three spin-weighted spherical harmonics~\eqref{eq:SpHTripleInt}, and the reflection symmetry of the Wigner-3j symbol~\eqref{eq:Wigner3j_reflection}, we get
\be
\begin{aligned}
\label{eq:YlmRlm2Int_1}
\int_{S^2}{\rm d}^2\hatOm
   \,&Y_{LM}(\hatOm)\,
   \sum_A F_{l m}^A(\hatOm)\,F_{l' m'}^{A*}(\hatOm) 
   \\
   &=8\pi^2(-1)^{l+l'+m}\sqrt{\frac{(l-2)!(l'-2)!}{(l+2)!(l'+2)!}}\sum_{s=-2,2} \int_{S^2}{\rm d}^2\hatOm
   \,Y_{LM}(\hatOm)\,{}_{-s}Y_{l, -m}(\hatOm)\,{}_{s}Y_{l' m'}(\hatOm)\\
&=8\pi^2(-1)^{l+l'+m}\sqrt{\frac{(l-2)!(l'-2)!}{(l+2)!(l'+2)!}}\sqrt{\frac{(2l+1)(2l'+1)(2L+1)}{4\pi}}\begin{pmatrix}
l & l' & L\\
-m & m' & M
\end{pmatrix}\\
&
\hspace{0.5in}\times
\left[1+(-1)^{l+l'+L}\right]\begin{pmatrix}
l & l' & L\\
2 & -2 & 0
\end{pmatrix}\,.
\end{aligned}
\ee
Recall that the Wigner-3j symbol vanishes unless $-m+m'+M=0$ (which implies $M=m-m'$) and $|l-l'|<L<l+l'$.

A useful special case of the previous equation for $L=0$, $M=0$ is
\be
\label{e:simple-int}
\int_{S^2}{\rm d}^2\hatOm
   \sum_A F_{l m}^A(\hatOm)\,F_{l' m'}^{A*}(\hatOm)
=(4\pi)^2 \delta_{ll'}\delta_{mm'}\,\frac{(l-2)!}{(l+2)!}\,.
\ee
Using this equation and expressions for $F^A(\hatOm, \hatp)$ and $\mu_{AA'}(\gamma,\hatOm,\hatOm')$ in terms of $F^A_{l m}(\hatOm)$ given in Sec.~\ref{s:Flm}, one can show that 
\begin{align}
&\mu_{\rm u}(\gamma) \equiv
\int \frac{{\rm d}^2\hatOm}{4\pi}\> \sum_A F^A_1(\hatOm)F^A_2(\hatOm)
=\sum_{l\ge 2} a_l P_l(\cos\gamma)\,,
\label{e:hd-harmonic}\\
&\widetilde{\mu^2}(\gamma)
\equiv 
\int\frac{{\rm d}^2\hatOm}{4\pi}
\int\frac{{\rm d}^2\hatOm'}{4\pi}\sum_A\sum_{A'} \mu_{AA'}^2(\gamma,\hatOm,\hatOm')
=\sum_{l\ge 2}\frac{a_l^2}{2l+1}
P_l^2(\cos\gamma)\,,
\label{e:tildemu2-harmonic}
\end{align}
where $a_l$ are given in \eqref{e:a_ell}.
The above two quantities appear in the expressions for the mean and cosmic variance for the ensembles discussed in this paper.

\bibliography{ref}

\end{document}